\documentclass[preprint]{aastex}

\newcommand{\pow}{$^{-1}$}
\newcommand{\poww}{$^{-2}$}
\newcommand{\oxy}{[O~$\!${\footnotesize III}]}
\newcommand{\degg}{$^{\rm o}$}
\newcommand{\kms}{\ km s$^{-1}$}
\newcommand{\resolp}{$\Delta\lambda/\lambda$}
\newcommand{\galaxy}{NGC~4151}
\newcommand{\gal}{NGC~1068}
\newcommand{\ltsim}{\raisebox{-.5ex}{$\;\stackrel{<}{\sim}\;$}}

\newcommand{\HST}{\textsl{HST}}
\newcommand{\angstrom}{\AA}
\newcommand{\fig}{Figure~}

\shortauthors{Das et al.}  \shorttitle{Kinematics of the NLR in
NGC~4151}

\begin{document}

\title{Mapping the Kinematics of the Narrow-Line Region in the Seyfert Galaxy
  \galaxy\altaffilmark{1}}

\author{V. Das\altaffilmark{2}, D.M. Crenshaw\altaffilmark{2,3},
  J.B. Hutchings\altaffilmark{4}, R.P. Deo\altaffilmark{2,5},
  S.B. Kraemer\altaffilmark{6}, T.R. Gull\altaffilmark{7},
  M.E. Kaiser\altaffilmark{8}, C.H. Nelson\altaffilmark{9},
  and D. Weistrop\altaffilmark{10} }

\altaffiltext{1}{Based on observations made with the NASA/ESA Hubble
  Space Telescope. STScI is operated bt the Association of Universities
  for Research in Astronomy, Inc., under NASA contract NAS5-26555. These
  observations are associated with proposal GTO-8473}

\altaffiltext{2}{Department of Physics and Astronomy, Georgia State
  University, Astronomy Offices, One Park Place South SE, Suite 700,
  Atlanta, GA 30303, das@chara.gsu.edu }

\altaffiltext{3}{crenshaw@chara.gsu.edu}

\altaffiltext{4}{Dominion Astrophysical Observatory, National Research
  Council of Canada, Herzberg Institute of Astrophysics, 5071 West
  Saanich Rd., Victoria, BC V9E 2E7, Canada,
  john.hutchings@nrc-cnrc.gc.ca}

\altaffiltext{5}{deo@chara.gsu.edu}

\altaffiltext{6}{Catholic University of America and Laboratory for
  Astronomy and Solar Physics, NASA's Goddard Space Flight Center, Code
  681, Greenbelt, MD 20771, stiskraemer@yancey.gsfc.nasa.gov.}

\altaffiltext{7}{Laboratory for Astronomy and Solar Physics, NASA's
  Goddard Space Flight Center, Code 681, Greenbelt, MD 20771,
  theodore.r.gull@nasa.gov}

\altaffiltext{8}{Department of Physics and Astronomy, John Hopkins
  University, 3400 North Charles Street, Baltimore, MD 21218-2695,
  kaiser@pha.jhu.edu}

\altaffiltext{9}{Department of Physics and Astronomy, Drake
  University, Des Moines, IA 50311, charles.nelson@drake.edu}

\altaffiltext{10}{Department of Physics, University of Nevada, Las
  Vegas, 4505 Maryland Pkwy., Las Vegas, NV 89154-4002,
  weistrop@physics.unlv.edu}

\begin{abstract}
  Using \emph{The Hubble Space Telescope's} Space Telescope
  Imaging Spectrograph (\emph{HST's STIS}), observations of the \oxy\ emission from
  the narrow-line region (NLR) of \galaxy\ were obtained and radial
  velocities determined. Five orbits of \HST\ time were used to obtain
  spectra at five parallel
  slit configurations, at a position angle of 58\degg, with spatial
  resolution 0\arcsec$\!$.2 across and 0\arcsec$\!$.1 along each slit. A
  spectral resolving power \mbox{(\resolp)}\ of $ \sim $9,000 with the
  G430M grating gave velocity measurements accurate to $ \sim
  $34\kms. A kinematic model was generated to match the radial
  velocities, for comparison to previous kinematic models of biconical
  radial outflow developed for low-dispersion spectra at two slit
  positions. The new high-resolution spectra permit the measurement of
  accurate velocity dispersions for each radial-velocity component. The
  full-width at half-maximum (FWHM) reaches a maximum of
  \mbox{1000\kms}\ near the nucleus, and generally decreases with
  increasing distance to about 100\kms\ in the extended narrow-line
  region (ENLR), starting at about 6\arcsec\ from the nucleus. In
  addition to the bright emission knots, which generally fit our model,
  there are faint high velocity clouds which do not fit the biconical
  outflow pattern of our kinematic model. These faint clouds occur at
  the turnover points of the outflowing bright clouds. We suggest possible
  scenarios that could explain these rogue clouds: (1) backflow resulting
  from shocks and (2) outflow outside of the bicones, although the latter
  does not explain how the knots are ionized and accelerated. A
  comparison of our observations with a high-resolution radio map shows
  that there is no evidence that the kinematics of the NLR clouds are
  affected by the radio lobes that comprise the inner jet.
\end{abstract}

 ~~~~~

\section{Introduction}
Seyfert galaxies are generally classified into two groups, based on
their spectra and emission line widths in the optical. Seyfert 1 galaxies are
characterized by broad permitted lines with typical FWHM velocities
$\geq$\,1000\kms, and narrower forbidden lines with FWHM
$\approx$\,500\kms, with a non-stellar underlying continua.
The spectra of Seyfert 2 galaxies show only the
narrow emission lines, and their non-stellar contribution is normally
detected only in polarized light. The generally accepted unification scheme
states that the two classes of galaxies are not disparate, but that
Seyfert 2 galaxies are Seyfert 1s with nuclei that are obscured by a dusty
molecular torus. \citet{AntMiller} discovered broad Balmer and
Fe$\,${\footnotesize II} lines in the Seyfert 2 galaxy\ \gal. From
polarization arguments they concluded that the lines must be emanating
from regions synonymous with the broad-line region (BLR) of Seyfert
1s. Subsequent studies reinforced the notion that Seyfert 2 BLR are
obscured from the line of sight but can be observed via scattered
radiation \citep{Ant}.  Therefore it is now generally accepted that
the Seyfert designation 1 or 2 is contingent upon the orientation
angle as perceived by the observer.\\
\indent A more dynamical
analysis is needed to determine if these two
types of active galactic nuclei (AGN) are indeed similar and powered by the same
processes. Kinematic studies of the narrow line region (NLR) can provide insights into the
intricate dynamical forces occurring within its vicinity. With the STIS spectrograph
aboard \HST, such studies can be easily accomplished efficiently. High
spatial and spectral resolution data in particular can provide us with
detailed diagnostics of the NLR and its vicinity. We have therefore
undertaken an investigation of the kinematics across the NLR of
\galaxy, one of the nearest Seyfert galaxies.\\
\indent This type of
observation and kinematic modelling was previously done on \gal, a bright Seyfert~2
galaxy \citep{CrenKrae}. The data were taken with the low resolution
(\resolp\ = 1000) G430L grating of STIS, and the radial
velocity measurements were made using \oxy\ $\lambda$~5007 emission
along two slit positions 0\arcsec$\!$.1 wide. They found that a simple
biconical outflow model closely matched the observed radial velocities
such that emission-line knots accelerate out from the nucleus to $ \sim $ 1000
\kms\ then decelerate back to the systemic velocity. The same model
was applied to \galaxy\ \citep[hereafter Paper 1]{CrenKrae2}, based on
STIS long-slit low resolution and slitless medium resolution spectra
(Hutchings et al. 1999; Kaiser et al. 2000; Nelson et al. 2000). In
Paper 1 they
pointed out that the observed velocity field can be well matched by
tilting the model bicone by about 40\degg\ out of the plane of the
sky, as opposed to only 5\degg\ for \gal.\\
\indent In this paper, we
present the same modeling procedures as in the two previous studies, but
at a spectral resolving power of 9000, and five parallel slit
positions covering the NLR of \galaxy. This provides the most detailed
map of the kinematics in a Seyfert 1 NLR, similar to that obtained for
the Seyfert 2 galaxy \gal\ \citep{Cecil}. We used a systemic velocity of
997\kms, based on H {\footnotesize I} observations of the outer
regions of the host galaxy \citep{Pedlar} and a Hubble constant of
75\kms\ Mpc\pow, which gives us a distance of 13.3\,Mpc to \galaxy, with
0\arcsec$\!$.1 corresponding to a projected linear scale of 6.4\,pc on
the sky.

\section{Observations and Analysis}
The STIS observations were taken on 2000 July 02 over five orbits
(proposal ID 8473, PI: J. Hutchings) with the \HST/STIS CCD through a
slit of \mbox{52\arcsec$\times$\,0\arcsec$\!$.2}, and dispersed with
the medium-resolution G430M grating, whose spectral resolving power is
$ \sim $9000. The STIS image scale is
0\arcsec$\!$.05/pix and the spatial resolution is 0\arcsec$\!$.1/pix in the cross-dispersion
direction. The five orbits yielded spectra for five parallel slit positions, each
at a position angle of 57.8\degg. There were three exposures per slit
with a total of 40 minutes of integration for slit 1 and 48 minutes for the remaining
slits. \fig\ref{intensity} shows the slit
placements and orientation superimposed on a WFPC2 \oxy\ emission-line
image of \galaxy\ showing the inner 4\arcsec.
Slit 1 was
centered on the optical continuum nucleus and the remaining slits were placed at
0\arcsec$\!$.2 and 0\arcsec$\!$.4 on either side, parallel to slit
1. The image was taken with the \oxy\ filter, which has a central
wavelength of 4961, and a bandpass of
4820--5100 \angstrom, which includes the H$
{ \beta } $ line and the \oxy\ $ \lambda \lambda $ 4959, 5007
redshifted emission lines. Most of the emission comes from within
2\arcsec\ of the bright central knot, although \oxy\ emission can be
traced as far as 6\arcsec\ on either side.\\
\indent
\fig\ref{image-OIII-slit1} shows a fully reduced STIS image of the two \oxy\
lines as they appeared through slit 1. The bright horizontal line
through the image is the continuum emission produced by the
nucleus. The top of the image corresponds to the northeast direction
and the horizontal scale is in wavelength. Both \oxy\ lines clearly
show an overall trend in that the majority of clouds in the upper region are receding
or redshifted and those clouds in the bottom part of the image are
approaching or blueshifted. Close to the nucleus, within $ \sim $
2\arcsec, the clouds appear highly dispersed.\\
\indent We used the IDL software
developed at NASA's Goddard Space Flight Center for the STIS
Instrument Definition Team to reduce the spectra. Cosmic-ray hits were identified and removed from
observations by comparing the three images obtained during each
orbit. Hot or warm pixels were replaced by interpolation in the
dispersion direction. Wavelength calibration exposures obtained after
each observation were used to correct the wavelength scale for
zero-point shift. The spectra were geometrically rectified and
flux-calibrated to produce a constant wavelength along each slit
column and fluxes in units of ergs s\pow\,cm\poww\,\angstrom\pow\ per cross
dispersion pixel.\\
\indent Each spectral image produced one spectrum
per cross-dispersion pixel along the slit, and centered
around the two bright doppler-shifted \oxy\ emission lines.  We used
the brighter of the two emission lines ($\lambda$5007) and fitted each
\oxy\ component with a local continuum and Gaussians. Noisy
spectra roughly $\geq$ 6\arcsec\ from either side of the nucleus were
not fit. Close to the nucleus, and on either side, emission knots
showed two and sometimes three major kinematic components
(\fig\ref{gaussfits}), and in such cases we fitted each identifiable
peak with Gaussians.\\
\indent \fig\ref{gaussfits} depicts a
progression of spectra illustrating how the components
fluctuate with position.
The graphs represent spectra taken from slit
1 and range in distance from 0\arcsec$\!$.3 to 0\arcsec$\!$.55 in
increments of 0\arcsec$\!$.05, going away from the nucleus in the
northeast direction.  In (a), there are two distinct kinematic
components of emission, which we fit with two Gaussians. Going from (a) to (b), a third component
emerges. A possible fourth component emerges in (c), but was not
measured
by our fitting routine. The second component shrinks in (c), and
disappears in (d), while the fourth component is now detected and
labeled `2'. This new second component grows stronger while the third shrinks and disappears
from (d) through (f). The appearance of the same components in
multiple rows along the slit indicates that we are obtaining multiple
measurements of spatially-resolved knots, the brightest of which can
be seen in \fig\ref{intensity}. For comparison purposes, we separated these
components by relative intensities and plotted them in different
colors (later in the paper).\\
\indent The central peaks of the Gaussians give us the central
wavelengths from which we used the Doppler formula to find radial velocities for each
component. There are three sources of uncertainty in the velocity
measurements. The first results from the fact that the \oxy\
emission lines are not Gaussians as shown in \fig\ref{gaussfits}. By
actually measuring the centroid
of a line for a number a spectra, we found that our Gaussian fitting
routine introduced a maximum fitting error of 0.75~\angstrom\ in the centroid. The second
error comes from emission cloud displacements from the center of the 0\arcsec$\!$.2
slit in the dispersion direction. We calculated that a 0\arcsec$\!$.1 offset would contribute an error
of 0.56~\angstrom. Noisy spectra introduced the third error. Various noisy
spectra was measured several times and the wavelength shifts were
found to fluctuate by 0.30 \angstrom\ maximum.
The errors were converted to velocities
and added in quadrature to produce a total maximum error of $\pm$
60\kms. We also varied the continuum placement for randomly selected
spectra but found that this did not shift the central wavelength
significantly.\\
\indent The S/N ratio of the integrated \oxy\ flux was large near the nucleus, approaching values
of 1000, and then decreased rapidly to $ \sim  $100 at about 1\arcsec\
from the nucleus. The S/N ratio reduced
to $ \sim  $ 50 at 3\arcsec, $ \sim  $30 at 4\arcsec, and $ \ltsim  $10
at 6\arcsec. We did not attempt to measure the emission if the S/N ratio
was $ <  $3, which typically occured at distances of $ >
$6\arcsec. This trend was seen in all five slits.\\
\indent In Paper 1, mostly bright clouds with one or two kinematic
components were detected and fitted to a model. In this paper, our data show
multiple components (1, 2, and 3) and hence in order to make a
comparison to Paper 1, we separated these kinematic components
according to total flux in the line. We assigned the color red to
components with the strongest flux; blue, to components with the the
next strongest flux, and black, to components with the weakest flux.
This kind of separation of the components
by fluxes incorporates a minor
problem. Consider \fig\ref{cartoon-spectra} showing a progression of
spectra from (a)--(f). The components are ordered 1, 2, and 3 going
from left to right, and the colors that we assign are marked below each
component. Notice that in (b), the second component is labeled
black (since it has less flux under the line) and the
third component is labeled blue. For similar reasons, the black and
blue colors switched places again in (c), as a new second component
emerged. The third component shrinked in (d), and disappeared in
(e). The relative nature of our separation criteria is made clear in
(e)--(f); with only two components detected, they
are labeled red and blue, and when we have detected only one
component, we labeled those red. In our velocity plots, this results in
occasional mixing between the low and medium flux clouds, but does not
significantly affect our model, since it was based on a general
fit for all points. The plot of fluxes (bottom plot) in
Figures \ref{slit1-vel-fwhm-flux} and \ref{slit-2to5-vel-fwhm-flux} clearly
show the effects of misidentification are small; as one can see the colors are clearly separated.\\
\indent Figures \ref{slit1-vel-fwhm-flux} and
\ref{slit-2to5-vel-fwhm-flux} show the kinematic components in plots
of radial velocities, FWHM, and fluxes as a function of projected position from
the nucleus. Negative positions corresponds to the southwest
direction. Note that the points tend to clump together, which reflects the
ability of STIS to resolve the clouds. In the upper plots, the brightest emission line
clouds at each position in
the northeast are red shifted, while those in the southwest are
blue shifted relative to the host galaxy. The highest velocities
occur closer to the nucleus and velocities as high as 800\kms\ can be seen in
approach and recession. The general trend is an increase in radial
velocity from near zero at the nucleus to a maximum redshift or
blueshift within $ \sim  $1\arcsec\ of the nucleus. Between $ \sim  $2\arcsec\ to
$ \sim  $4\arcsec\ the velocities reduce from about
200\kms\ to 100\kms\ and at distances of about 4\arcsec\ or more from
the nucleus, the radial velocities taper off to the ENLR rotational
velocities. The medium flux clouds
generally show the same pattern as the bright ones except in slits 1
and 4 where they are not confined to only blue shifts in the SW.
Also in slit 2, the
medium flux clouds show blueshifts in the NE. The low flux clouds
also follow similar patterns as the bright clouds except in slits 1 and
4 where one can see additional flow patterns opposite to that of the bright
clouds. The FWHM in the middle plots
show a range of widths at each position, but a decreasing average
width with distance from the nucleus at all
flux levels. In the bottom plot, all three components show a
strong decrease in flux with distance and significant structure that
corresponds to the emission line knots.\\
\indent \citet{Hutchings} found a number of low-flux, high velocity
clouds in \galaxy\ using a combination of WFPC2 and STIS images and
STIS slitless spectra. Some of these faint clouds did not fit into the
general picture of biconical outflow (e.g. redshifts were detected
in the SW where blueshifts dominate). To check for consistency, we
identified Hutchings et al.'s clouds in our data and compared our
velocities with theirs. Our measurements are closely
correlated over a wide range of velocites as shown in
\fig\ref{Hutchings-Das-Velocity}. Also shown on
top of the figure are the letters of various high velocity knots,
labeled according to Hutchings' scheme. The positions
of Knots E and J in Hutchings et al.'s image should have placed them in our slits 1
and 5 respectively, but they were not found in our data. On
average, the difference between Hutchings et al.'s
velocity and ours is 134\kms, but there is no significant systemic offset. In addition to these knots, we
found many more high velocity low and medium emission knots. These can
be seen in the velocity plots of Figures \ref{slit1-vel-fwhm-flux}
and \ref{slit-2to5-vel-fwhm-flux}.\\

\section{Kinematic Models and Results}
In \citet{CrenKrae}, the radial outflow from the
Seyfert 2 galaxy, \gal\ was accounted for by a simple biconical radial
velocity law, $v = kr$, $r$ being the distance from the central
source and $ k  $ being a constant. They used STIS low-resolution long-slit spectra taken in the
\oxy\ emission regime. The model bicone was evacuated along its axis
up to an inner opening angle of 26\degg\ from which it was assumed to
have a filling factor of 1 up to an outer opening angle of 40\degg.
The gas within the filled region accelerates from rest close to the nucleus to about 1300\kms,
at a distance 140 pc away. From this point it decelerates back to
systemic velocity at a distance of about 300\,pc from the nucleus.
The deceleration of the gas followed a law with the velocity
decreasing linearly with distance, $v = v_{  max  } - k^{  '   }r$. The best-fit model used an
inclination of 5\degg\ (relative to the plane of the sky) for the
bicone axis. In Paper~1, the same
modeling procedures were used to map the kinematics of \galaxy. The
bicones were assumed to have identical geometrical and physical
properties and a filling factor of 1 within the cones and no absorption
of the \oxy\ photons. Interestingly, the velocity
laws $v = kr$ and $v = v_{  max  }-k^{  '  }r$ further out also worked in this model. The middle column of
Table \ref{compare-ngc4151} shows the parameters for the best fit
model from Paper~1.\\
\indent In this paper we test the model in Paper~1
using higher resolution data and more coverage of the NLR of
\galaxy. The kinematic modelling code used in Paper~1 was updated to
test different velocity laws, corresponding to radial outflow. Parameters for the bicone that were
allowed to vary in our code include the extent of each cone along its
axis ($z_{max}$), the minimum and maximum half opening angles
($\theta_{inner}$ and $\theta_{outer}$), the inclination of its axis
out of the plane of the sky ($i_{axis}$), the velocity turnover radius
($r_{t}$) (measured radially outward from the nucleus), and the maximum
velocity at the turnover point ($v_{max}$). All the above parameters
were assumed to be the same for both top (northeast) and bottom
(southwest) parts of the bicone. We tested various velocity laws for
the outflow and
found that two of them fit the data reasonably well. The first law was $v = kr$ and $ v = v_{
  max  }-k^{  '  }r  $ for the acceleration and deceleration phases of
the gas respectively. The second law was $v = k\sqrt{r}$ and $ v = v_{
  max  }-k^{  '  }\sqrt{r}  $ for the two phases. We referred to these
laws as the $r\ law$ and the $ \sqrt{r}\ law  $
respectively\footnote{Note that the latter corresponds to constant
  acceleration with radius and not the former, as stated in some of our earlier
  papers. The acceleration in the papers was proportional to radius.
  This can be seen by solving the simple differential equation $a =
  \frac{dv}{dt} =
  v\frac{ dv   }{ dr   } = r^{  n  }  $, where
  n = 1 for the $r\ law$, and n = 0 for the  $
  \sqrt{r}\ law  $.}. A constant velocity law and velocity
proportional to higher powers of radius did not work well. Motion
perpendicular to the bicone axis and infall do not work (see Paper~1).\\
\indent Our code generates a two-dimensional velocity map and samples
this map with slit positions, orientations, and widths that match
those of the observations. \fig\ref{models} shows our model velocity
field projected onto the sky. The purple color is the highest
negative velocity and the red is the highest positive velocity, with intermediate
velocities given in gradients. Next to the fields are velocity
extractions from slit 1 done to visually aid the reader. The
extraction consists of four parts, two from the back side of the
bicone, and two from the front side. The upper two graphs consists of
velocities extracted from the top and bottom parts respectively, of
the back side of the bicone. Similarly the lower two graphs represent
the front side of the bicone. In each graph, the relevant extracted section is in
shaded grey, in addition to the total extraction from slit 1, which
put things into perspective.
Radial velocities were extracted along
the five slits and plotted as a function of position in arc seconds.
Note that the model was tilted out of the plane of the sky so
that the extracted radial velocities
would best match the general trend of the data. The other bicone
parameters were then adjusted until the
best match between model and data was established. In this fit,
emphasis was placed on the high and medium flux points. The low flux
points in slits one and four indicate a flow pattern which in some
cases is opposite
to that determined from the model; for the rest of the slits however,
there is general agreement.\\
\indent \fig\ref{plotshade} shows all
five slits with the models in shaded grey and the observed data points
in colors. Zero position corresponds to the nucleus. These radial
velocity plots show that the northeast side is mostly redshifted, and
the southwest side is mostly blueshifted, except for some low
luminosity clouds (black points), and a few medium luminosity clouds
(blue points), which exhibit an unusual flow pattern. These are shown
by the arrows in \fig\ref{plotshade} and to some extent, between 0\arcsec\ and -2\arcsec\
in \fig\ref{plotshade} (e). Plots (a) through (e)
represent models with the $r\ law$ and correspond to
slits 1 through 5 respectively. Plot (f) shows a fit for slit
one with the $ \sqrt{r}\ law  $. A closer inspection
of plots (a) and (f) shows minuscule differences in the fit, with (a)
being slightly better. Minor tweaking of
the parameters would result in good agreement for the $ \sqrt{r}\ law
$ in slit 1, however the $r\ law$ tends to be more
consistent across all five slits.\\
\indent The final model parameters for
\galaxy\ are presented in Table~\ref{compare-ngc4151}. For comparison,
the fitting parameters from Paper~1 are also presented. Following the
deceleration trend of the knots in all five
slits, the model suggests that the knots return to systemic
velocity at a larger distance from the nucleus than previously found
in Paper~1. This extends the bicone longer than in Paper~1, to 400\,pc as opposed to
288\,pc. The turnover point is much closer in, 96\,pc compared to
162\,pc and our inner opening angle is 15\degg, which makes our model less
evacuated along the bicone axis. The high luminosity knots seem to
travel outward from the nucleus at lower inclination and was modeled
in Paper~1 as such; however, in this paper we also incorporate medium
and low luminosity knots into the plots and hence our match suggests a larger
inclination of 45\degg. The other parameters, inner and outer
opening angles, and maximun velocity of the gas, show only minor differences
between the two papers. The maximum velocity before turnover was increased in our
paper to account for the fast moving medium and low luminosity clouds seen in
\fig\ref{plotshade}. These clouds are accelerated
to high speeds close to the nucleus and they therefore play a major
role in modelling the turnover point, which we found to be $\sim$ 96 pc. If we had used only
the bright clouds (red points) then our turnover point would have been larger
and similar to that of Paper~1.\\
\indent \fig\ref{fwhm-vel} shows a plot of FWHM versus radial velocity
for all five slits combined. There seems to be no definite correlation between
the FWHM and the absolute value of the radial velocity. This indicates that the directions
of velocity dispersion are not preferentially along the direction of
the velocity vector. The FWHM is therefore not due to a purely radial
gradient. However, various low luminosity clouds
do show some trend. \fig\ref{vel-fwhm-low-flux} shows evidence
of a correlation for a combined plot of isolated groups of points shown by the arrows in
\fig\ref{plotshade}. These groups of points do not follow the general
trend of the bright clouds and they consist of only medium and
low flux points. This leads us to believe that these
particular knots are either preferentially disturbed in the direction
of the velocity vector, or that they exhibit turbulence in
proportion to their velocity. Note that the faintest clouds have
higher FWHM and radial velocity. We will look at these rogue clouds more
carefully later in the paper.\\
\indent \fig\ref{bicone} shows a model of \galaxy\ with the bicone as it should
appear in the sky to us, using our final model parameters. The
position angle of the Galaxy is 22\degg, with an inclination of
20\degg\ to the line of sight, the SW side being closer to us. The
position angle of the bicone is 60\degg, and its
inclination is 45\degg. The maximum half opening angle of the bicone
is shown here to be 33\degg. Note that our line of sight is outside of
the bicone, at an angle of 12\degg\ with respect to the nearest part
of the bicone. The $ \beta  $ angle, which is the angle
between the bicone axis and the normal to the galactic disk, is
36\degg.
Due to the bicone's large opening angle, it will intersect a sufficiently thick
galactic disk (e.g., a scale height of $\sim$ 100 pc). The NE bicone
will also intersect the NE galactic
disk due to symmetry. This intersection produces the observed
geometry of the ENLR as shown by \citet{Evans}.\\
\section{Interaction with Radio Jet}
The kinematics of the \oxy\ clouds in some Seyfert galaxies
have been interpreted as outflow away from (perpendicular to)
the jet axis. Several authors have suggested
close relationships between the radio jets and the optical emission
such that hot gas around the radio lobes expands and propels the NLR clouds to
high velocities (see Gallimore et al. 1996; Capetti et al. 1997b: Axon
et al. 1998). Although this scenario seemed plausible when applied to
\gal, it failed to explain the turnover velocity where emission knots
started to decelerate back to systemic velocity, at about 2\arcsec\
from the nucleus \citep{CrenKrae}. Also the clumpy nature of the radio knots could
not explain the orderly large-scale flow pattern of the radial velocities. The
axis outflow was discussed in
\citet{NelWei} in relation to \galaxy. They found that regardless of
inclination, the radial velocities should appear both red-shifted and
blue-shifted with equal magnitudes at each position along the axis. However \galaxy\ has
mostly blue-shifted clouds in the southwest region and mostly red-
shifted clouds in northeast of the nucleus. Therefore, radial outflow
matches the data much better.\\
\indent To test whether the radio knots are responsible for locally
disturbing the \oxy\ clouds, we generated plots from radio maps taken
by the Very Long Baseline Array plus the Very Large Array
\citep[see][and references therein]{Mundell} together with radial velocity
plots from our data, and compared them within the 2\arcsec\ range that the
radio data cover. We extracted radio emission along the same slit
positions as our 5 slit positions in \fig\ref{intensity}. The radio
fluxes were summed across each 0\arcsec$\!$.2 slit width and normalized to
the peak of the brightest radio knot, which was in slit 1. Radio flux maxima in slits 3
through 5 were only about one tenth the value as those in slits 1 and 2.\\
\indent \fig\ref{radio-slit-new}
shows our slits overlaid on top of the radio map. The radio jet is not
parallel with the \oxy\ axis, but lies close to one edge of the projected
bicone. Slits 1 and 2 contained the two brightest radio knots and the entire image spans
about 4\arcsec. \fig\ref{radio-slit-all} shows intensity plots of the
radio emission together with the our radial
velocities. Here we expect that bright radio knots would produce roughly
symmetric red and blue shifts at the positions of the knots if the axis outflow applies to
\galaxy. However we see no evidence for such a scenario in any
slit. \fig\ref{radio-slit-all} (b) shows some blue shifted clouds at
the position of the radio knot, but no corresponding red shifted
clouds are observed. We see some red-shifted material in the upper
bicone at 0\arcsec$\!$.5, but this is too
far away to be affected by the radio knot. Near the brightest radio
flux in (c) and (d), we see some blue and red shifted \oxy\ clouds
respectively, but again there are no symmetric counterparts for these
clouds. Similar plots of the velocity widths show no evidence for
larger FWHM at the positions of the radio lobes. We conclude that the
radio lobes have no discernible effects on the kinematics of the NLR clouds.

\section{Discussion}
We confirm our previous biconical outflow models \citep{CrenKrae2},
that were used to match the radial velocities of the
NLR in \galaxy, using higher velocity-resolution STIS spectra. In \citet{CrenKrae2}, the model was based on two slit
positions at different position angles. Here we use five parallel
slits, all at the same position angle, but different from that used in
\citet{CrenKrae2}. However, the previous model was found to be consistent
with ours, except for a few adjustments. The turnover point of the velocity (where clouds start to
decelerate) occurs much closer to the nucleus than the previous model
showed, and the point where the clouds velocity approaches systemic
velocity is further away. These two differences follow from
the fact that we were able to detect fainter, high-velocity knots
closer in to the nucleus. We note that some of the high-velocity low-flux components do not
match the model as shown by the arrows in \fig\ref{plotshade}. These
high-velocity clouds had been detected in
\citet{Hutchings} to have velocities of both approach and recession on
both sides of the nucleus. We see similar trends in
\fig\ref{plotshade} (a), (b), (d), and somewhat in (e). The rest of the low-flux clouds
are in agreement with the model.\\
\indent The sum of the half opening angle
and the inclination of the axis puts the near-sided
south-west cone at an angle of 14\degg\ with respect to the line of
sight for Paper~1 and 12\degg\ for
this paper. This means that we are still looking outside the bicone,
contrary to the Seyfert 1 classification, and general Seyfert
unification models. The most likely explanation for an unobscured view of
the nucleus is that the bicone edges, and therefore the torus edges,
are not sharp.\\
\indent The radio jet lobes do not seem to cause any disturbance to
the \oxy\ emission and are not directly responsible for the
acceleration of the gas, as is evident from
\fig\ref{radio-slit-all}. One should see both blue and red-shifts
roughly at the same position in these
plots, but there is no evidence of this. The same plots with FWHM also show no
correlation between position of the radio lobes and kinematics of the
\oxy\ knots. \\
\indent We propose possible scenarios to explain the faint
clouds that do not fit the biconical outflow pattern, such as in
\fig\ref{plotshade} (a).
One explanation is
that of backflow due to shocks, as a result of outflowing bright
clouds impacting on an ambient medium.
The low luminosity clouds could be flowing back along the
bicone with a velocity that is proportional to the outflow
velocity; that is, the higher the outflow velocity, the higher the
backflow velocity. The top arrow in
\fig\ref{plotshade} (a) points to rogue points
which may have resulted from fast outflowing gas in the near side of
the southwest bicone, with the low emission gas moving in the opposite
direction of general outflow, leaving a trail of clouds in the
direction of the nucleus; see location b in \fig\ref{backflow-cartoon}. Because the
near side of the cone has a higher projected
velocity, we see the rogue clouds trailing back along this
side. The bright clouds are accelerated in the first few tenths of an arcsec and
the rogue points mimic the same behaviour, but in the opposite
direction. The same argument holds for clouds pointed at by the second
arrow in \fig\ref{plotshade} (a). This backflow most likely comes
from the far side of the cone (d in \fig\ref{backflow-cartoon}), since
the velocities and position there agree with those of the rogue clouds.
The knots pointed at by the arrow in \fig\ref{plotshade} (b) can be
associated with backflows from the near side of the NE bicone where their
positions coincide with those of the bright clouds at location a in
\fig\ref{backflow-cartoon}. Clouds
pointed at by the lower arrow in
\fig\ref{plotshade} (d) can also be explained by backflow from the
near side of the NE bicone. The velocities and positions of the rogue
clouds are similar to those of the clouds in this region. The argument loses some
credibility, however, when we consider the upper
arrow in \fig\ref{plotshade} (d). These high velocity rogue knots
seemed to be geometrically associated with the near side region of the
SW bicone, (b in \fig\ref{backflow-cartoon}), but
there are no bright clouds in this region to contribute to the
backflow behaviour.\\
\indent There is another possibility which can explain the unusual flow
pattern of all the rogue clouds. The argument is that the rogue clouds
are moving in an entirely different geometry flowing away from the
nucleus as shown in \fig\ref{backflow-cartoon}, locations e, f, g, and
h. Note that
backflows from a, b, c, and d, are equivalent to
projected velocities from g, f, e, and h, respectively. This
geometrical equivalent of the backflows is but one of the ways in
which the projected velocities agree, since the geometry could be
positioned anywhere along the line of sight to yield the same projected velocities. We
adopt the geometry whereby clouds emanate from the central nucleus, as
this scenario seems to be the most plausible one. If that is the case
then the flows from locations e and f would be at a rather large angle
(near 90 degrees) with respect to the bicone axis. This would suggest
that clouds are flowing close to the plane of the torus, which
might explain their low flux. The problem with
this scenario is that there is no means
for the clouds to be ionized unless radiation is coming from holes in
the torus. It is also not clear how these clouds would be
accelerated without radiation or winds from the nucleus, as these
clouds do indeed attain high velocities.\\
\indent We prefer the first scenario, because the rogue clouds appear
at the positions of the turnover velocities of the bright clouds. This
supports the idea (Paper 1) that the bright clouds may be
plunging into an ambient medium at the turnover point which cause the
clouds to slow down and eventually return to zero velocity, at the
same time producing the faint clouds in backflow. The
second scenario seems less likely, because there is no other evidence
at present for a leaky torus.

\acknowledgments
We thank C. Mundel for providing the radio map of \galaxy\ used in
this study.\\
The observations for this research were obtained as part of a
NASA Guaranteed Time Observer allocation to the STIS Science Team.

\clearpage

\figcaption[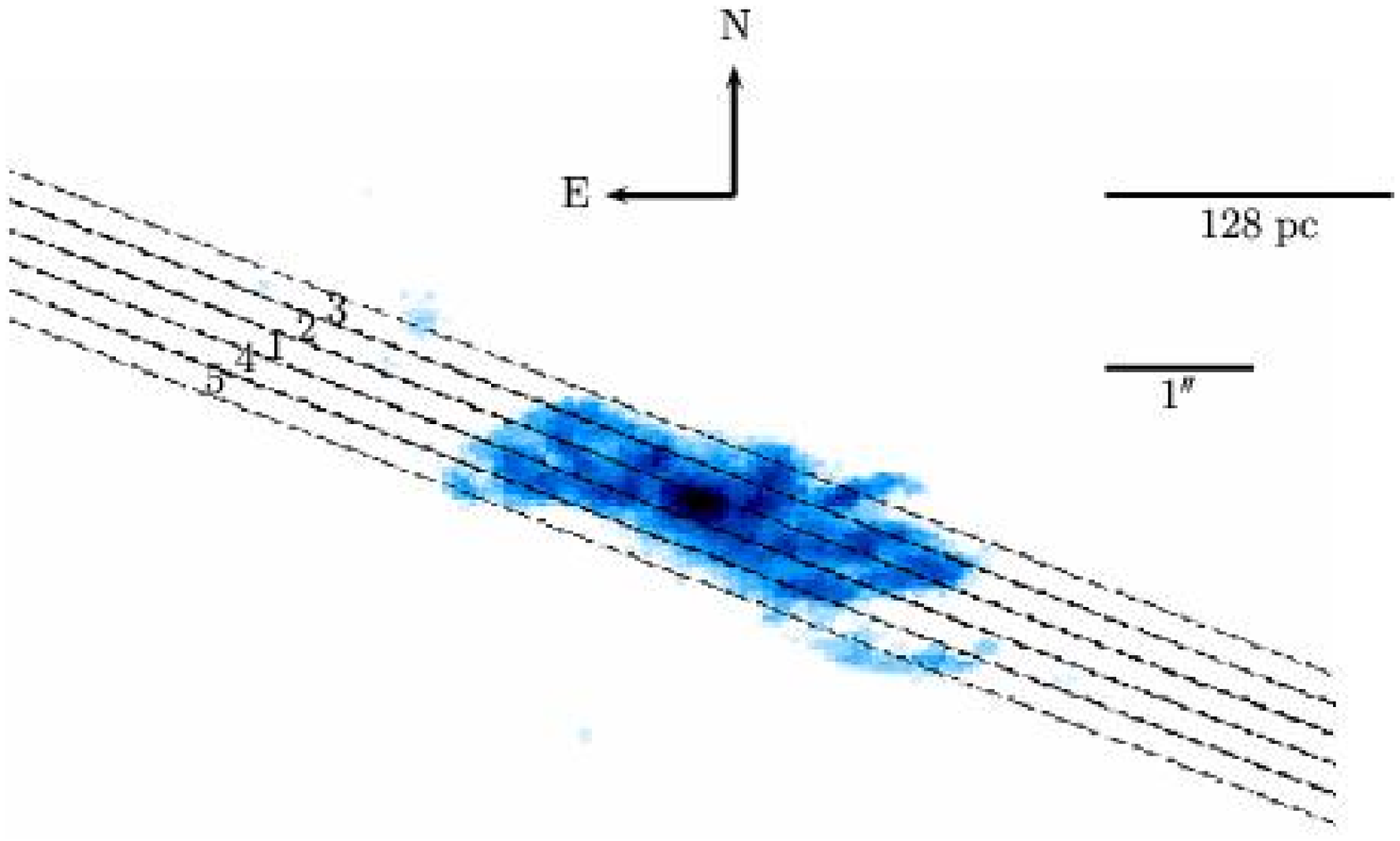]{\oxy\ intensity image of the NLR of
  \galaxy\ showing the orientation and position of the five slits
  taken by \HST\ STIS. The position angle of the slits on the sky is
  58\degg. \label{intensity}}

\figcaption[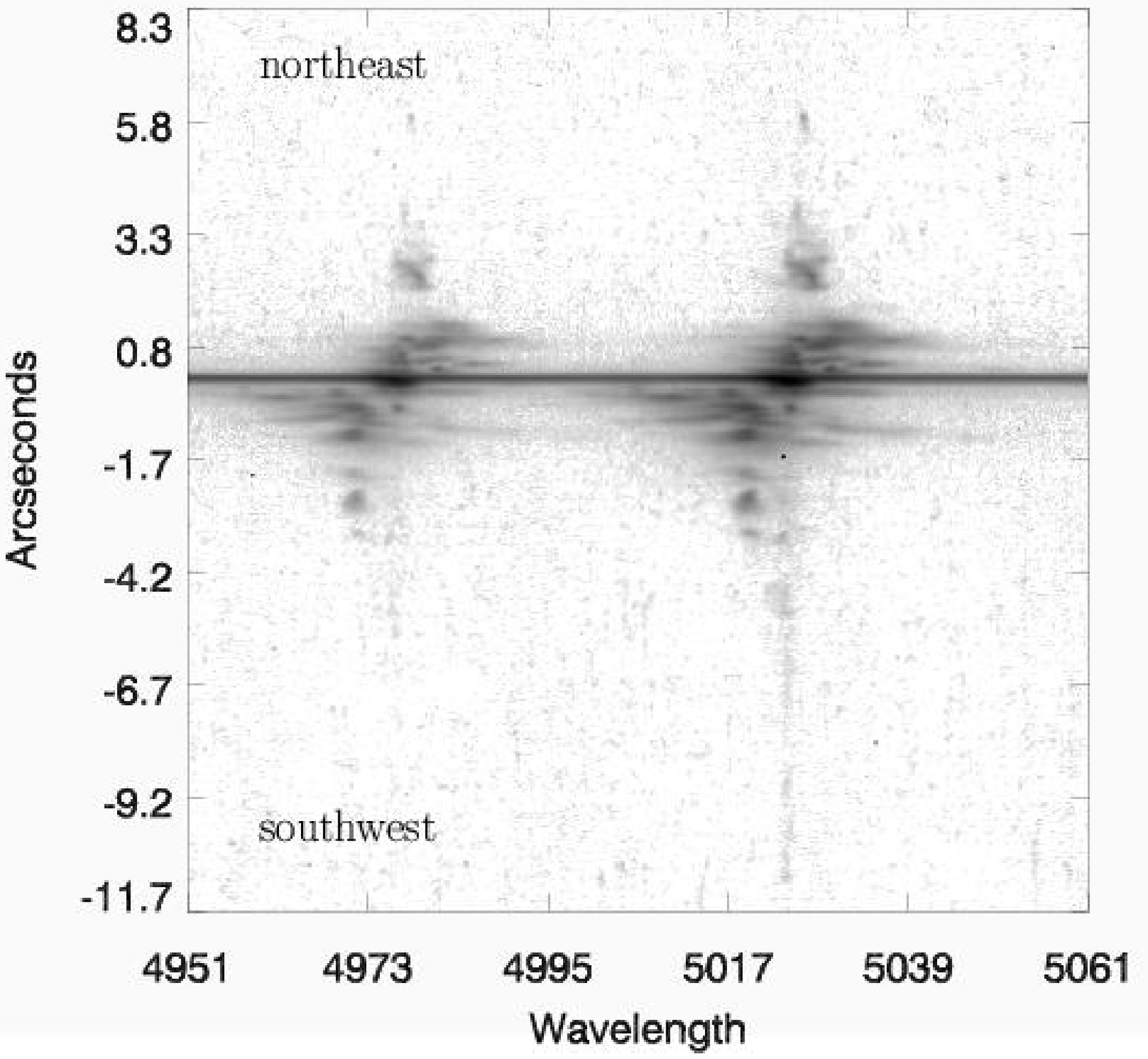]{STIS G430M fully
  reduced CCD spectral image for slit 1, showing the emission knots
  of \oxy\ $\lambda\lambda$ 4959, 5007.  The
  nucleus is the bright continuum band across the entire image. A couple of
  warm pixels have not been removed by our software. Spatial
  extent of the image (top to bottom) is 23.7\arcsec\ and the abscissa
  is in wavelength. \label{image-OIII-slit1}}

\figcaption[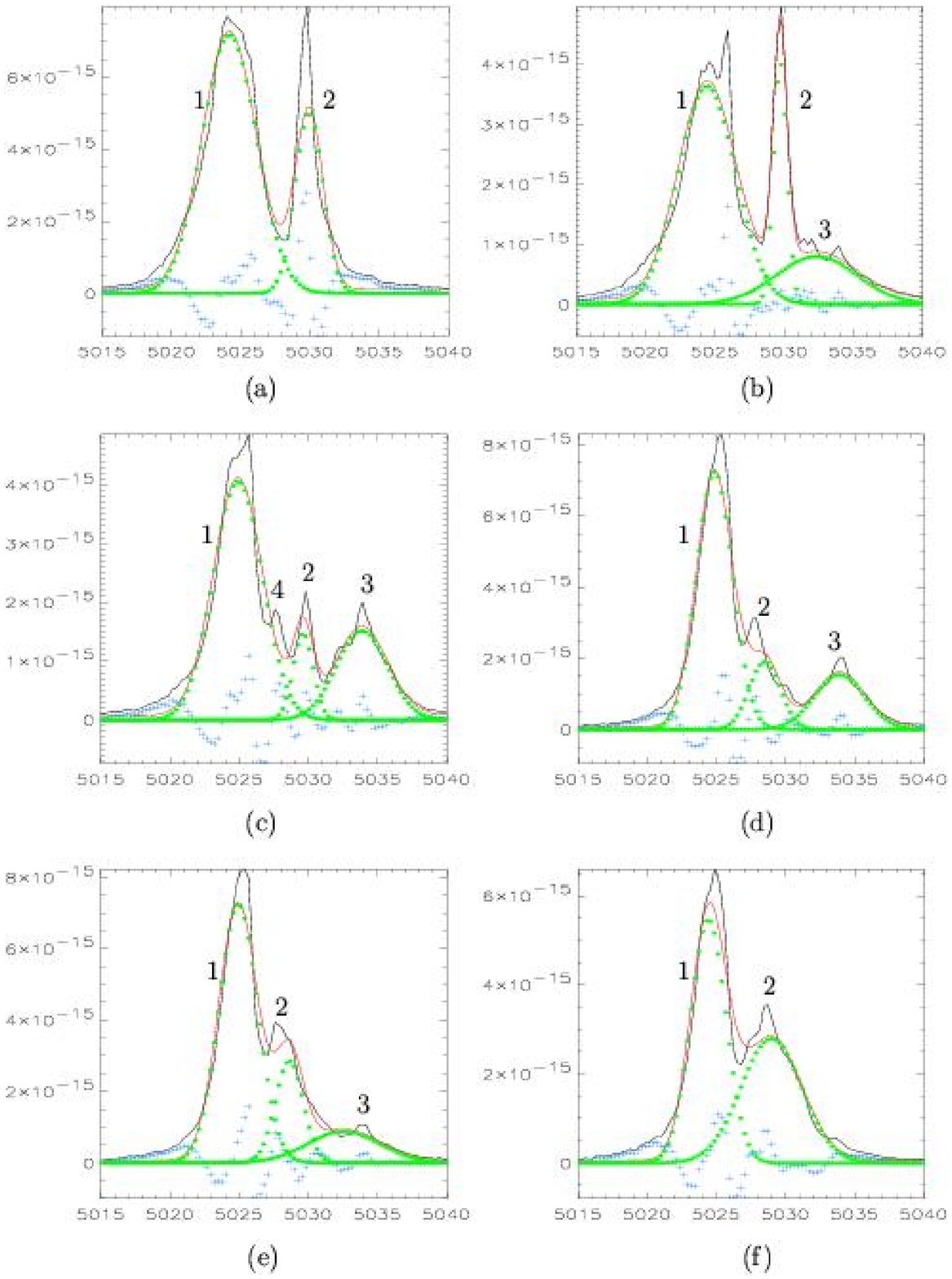]{Spectra in black and fits
  in color showing the multi-component gaussian fits. The plots,
  ordered (a)--(f) are
  from positions separated by 0\arcsec.05, increasing in distance from
  the nucleus. The green dotted curves represent the fit
  for each component, the red is the sum of all the green curves and
  the blue is the difference between spectrum, (black curve) and fitted
  red curve. The spectra were not yet doppler corrected for the
  receding motion of \galaxy. Wavelengths are in \angstrom\ on the abscissa and fluxes are
  in ergs s$^{-1}$ cm$^{-2}$ \angstrom$^{-1}$ on the ordinate. \label{gaussfits}}

\figcaption[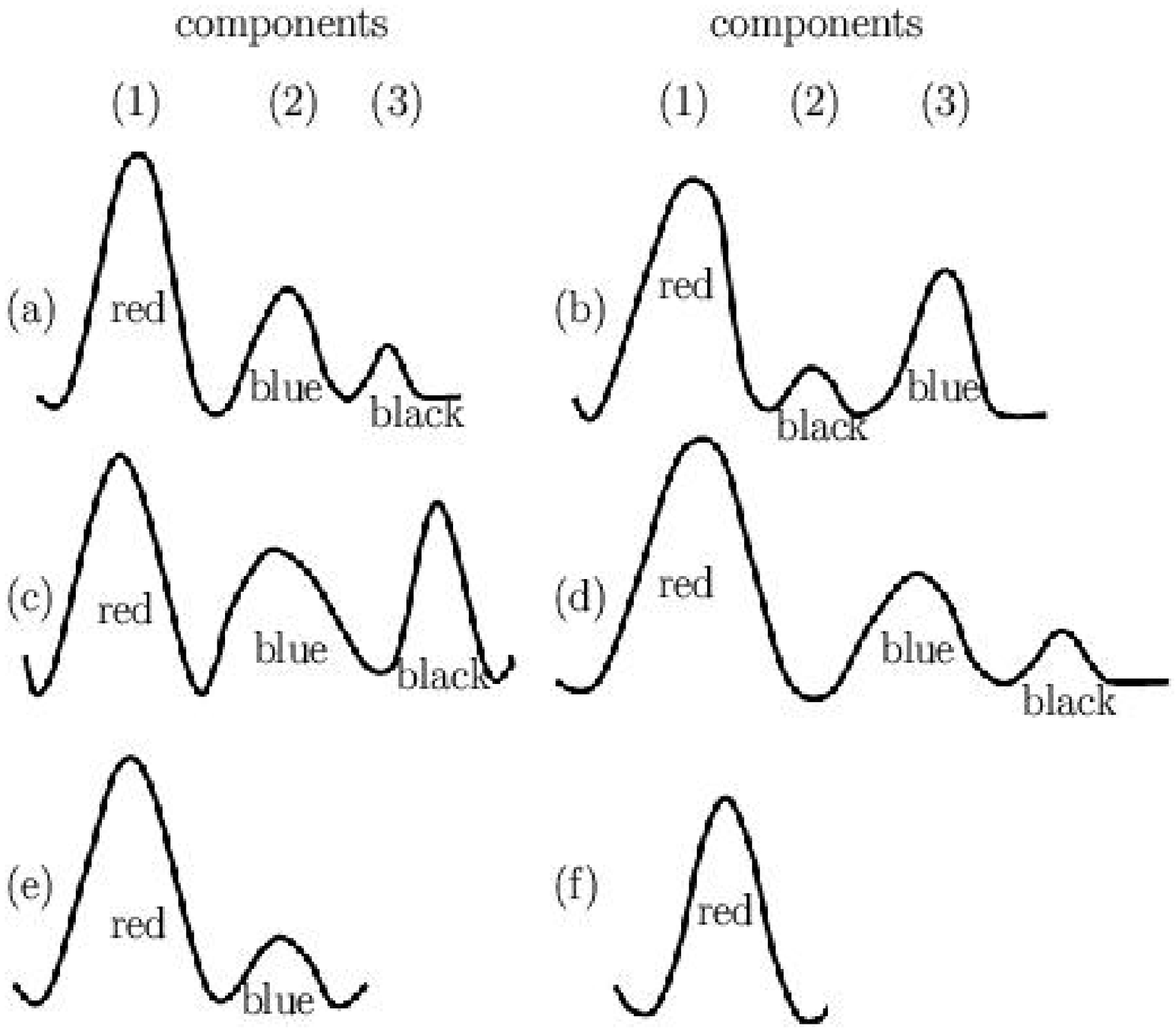]{Cartoon showing multiple
  components in a progression of spectra and how we separate them
  according to flux. The components are assigned the
  colors under them after separation. In (c) there is more flux under
  the line, hence the color blue is assigned. \label{cartoon-spectra}}

\figcaption[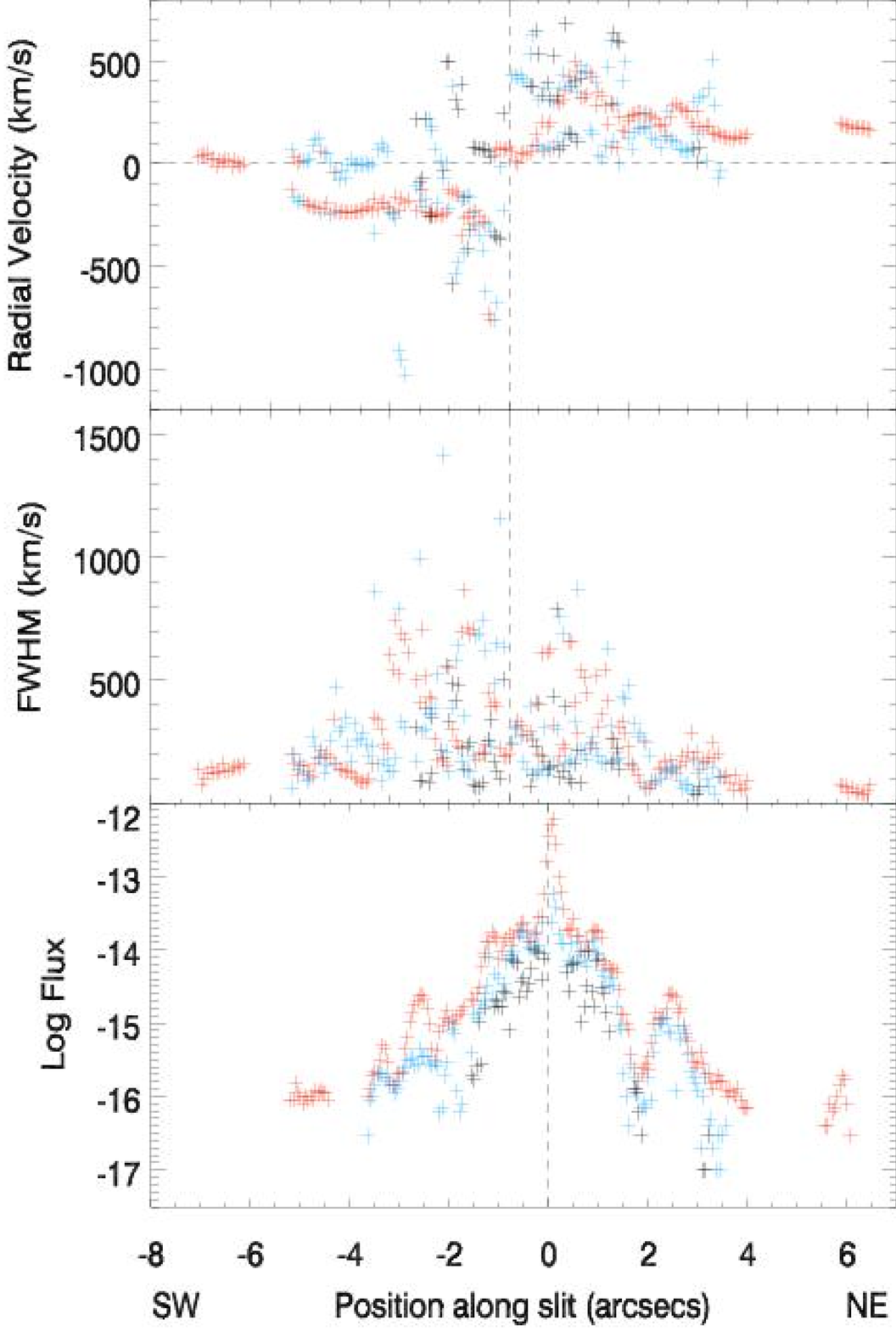]{Slit 1 multi-colored
  plots showing the components present in the emission
  clouds at each position. Measurements were made using the \oxy\ $\lambda$5007
  emission from the NLR of \galaxy. The colors represent high(red),
  medium(blue), and low(black) flux levels respectively. Radial
  velocities are given relative to systemic, and fluxes are in ergs
  s$^{-1}$cm$^{-2}$. All graphs share a common abscissa of position in arc-seconds
  (\arcsec) from the nucleus. The FWHM is not corrected for
  instrumental resolution. Note the rogue clouds in the upper left of
  the velocity plot. \label{slit1-vel-fwhm-flux} }

\figcaption[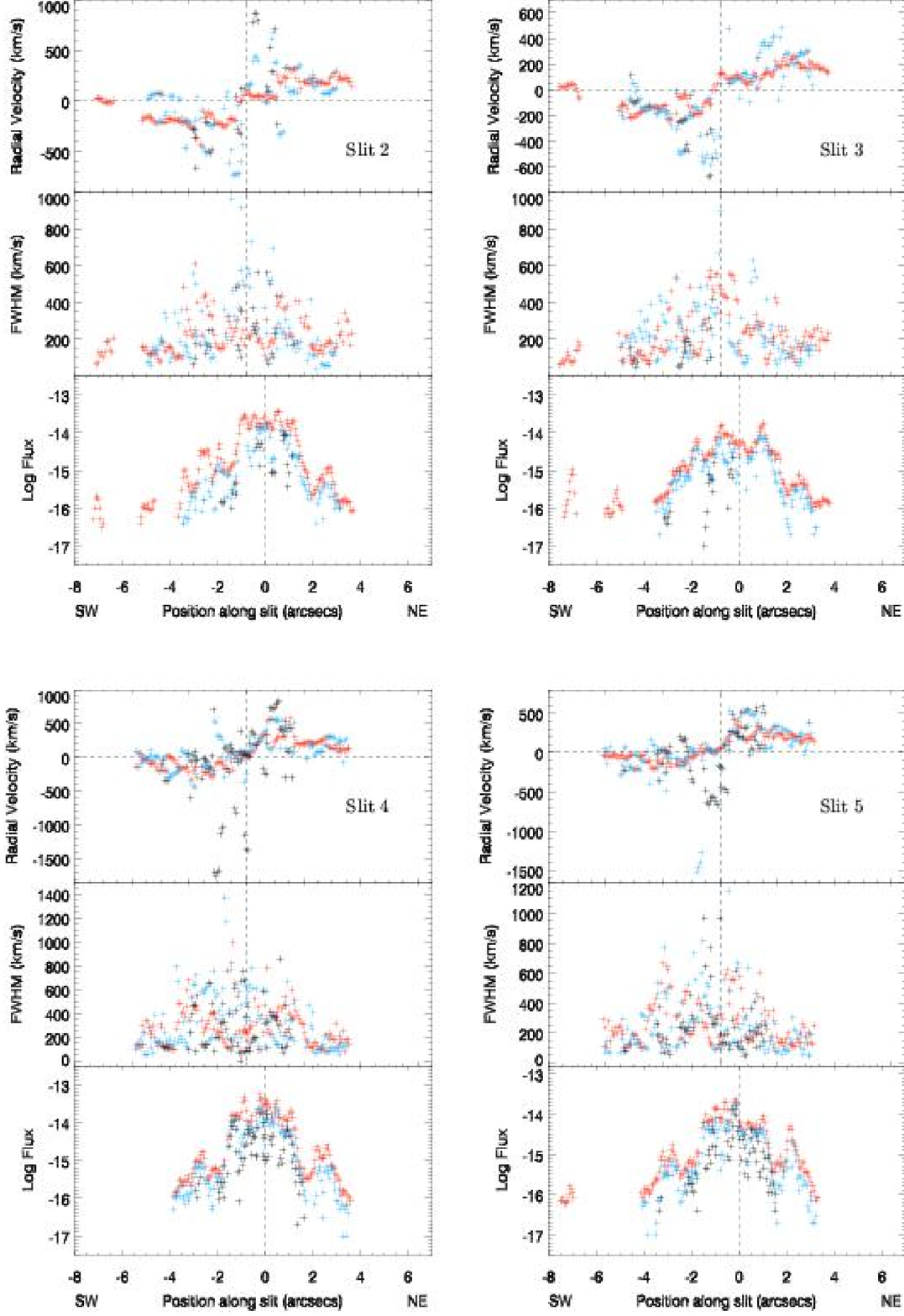]{Plots of velocity, FWHM, and flux for slits
  2--5. The colors depict the usual separation of flux levels as in
  \fig\ref{slit1-vel-fwhm-flux}.\label{slit-2to5-vel-fwhm-flux} }

\figcaption[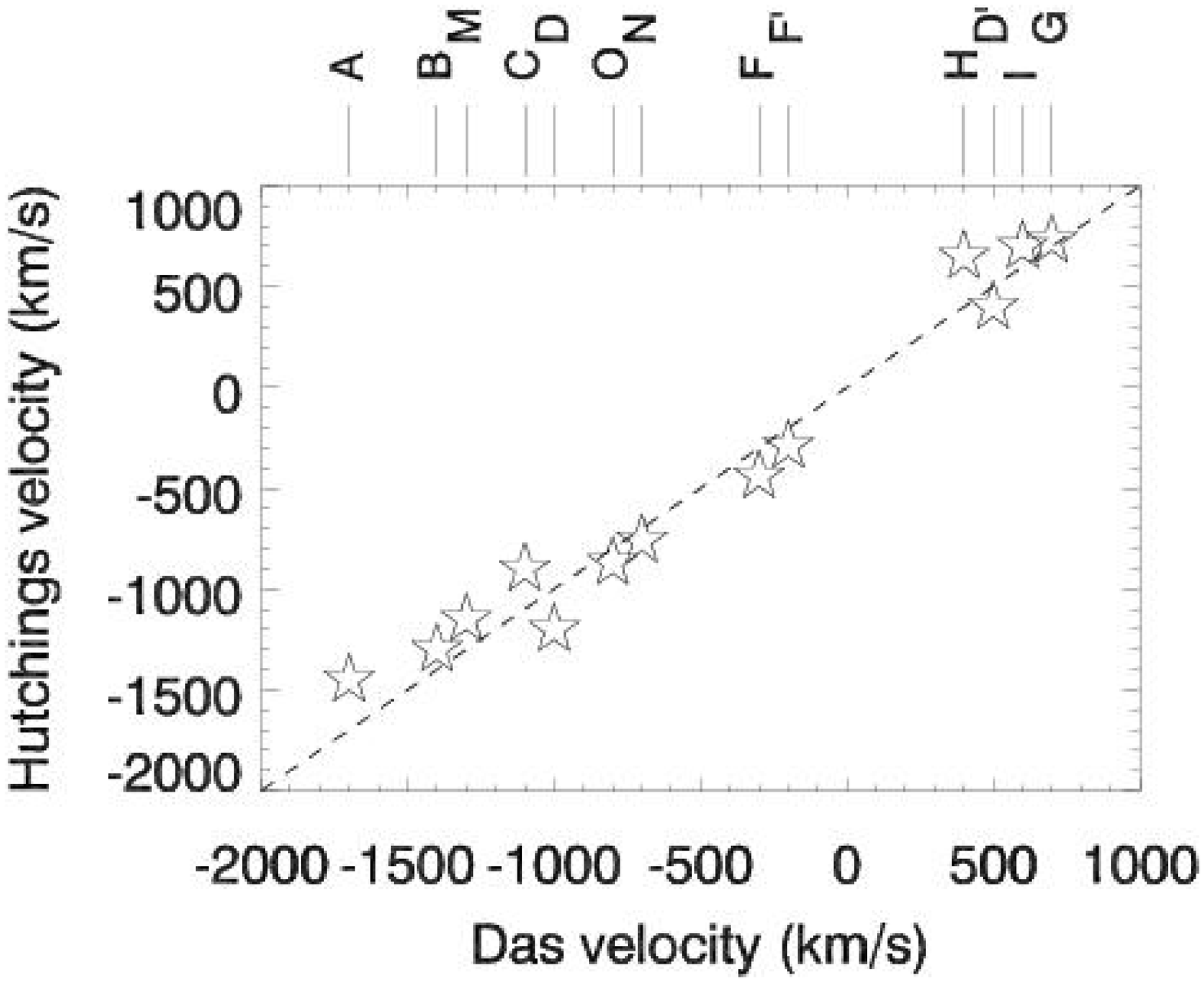]{Comparison between
  our and Hutchings et al.'s velocities. We found most of
  their knots, except knots E and J. See
  \mbox{\citet{Hutchings}.} \label{Hutchings-Das-Velocity} }

\figcaption[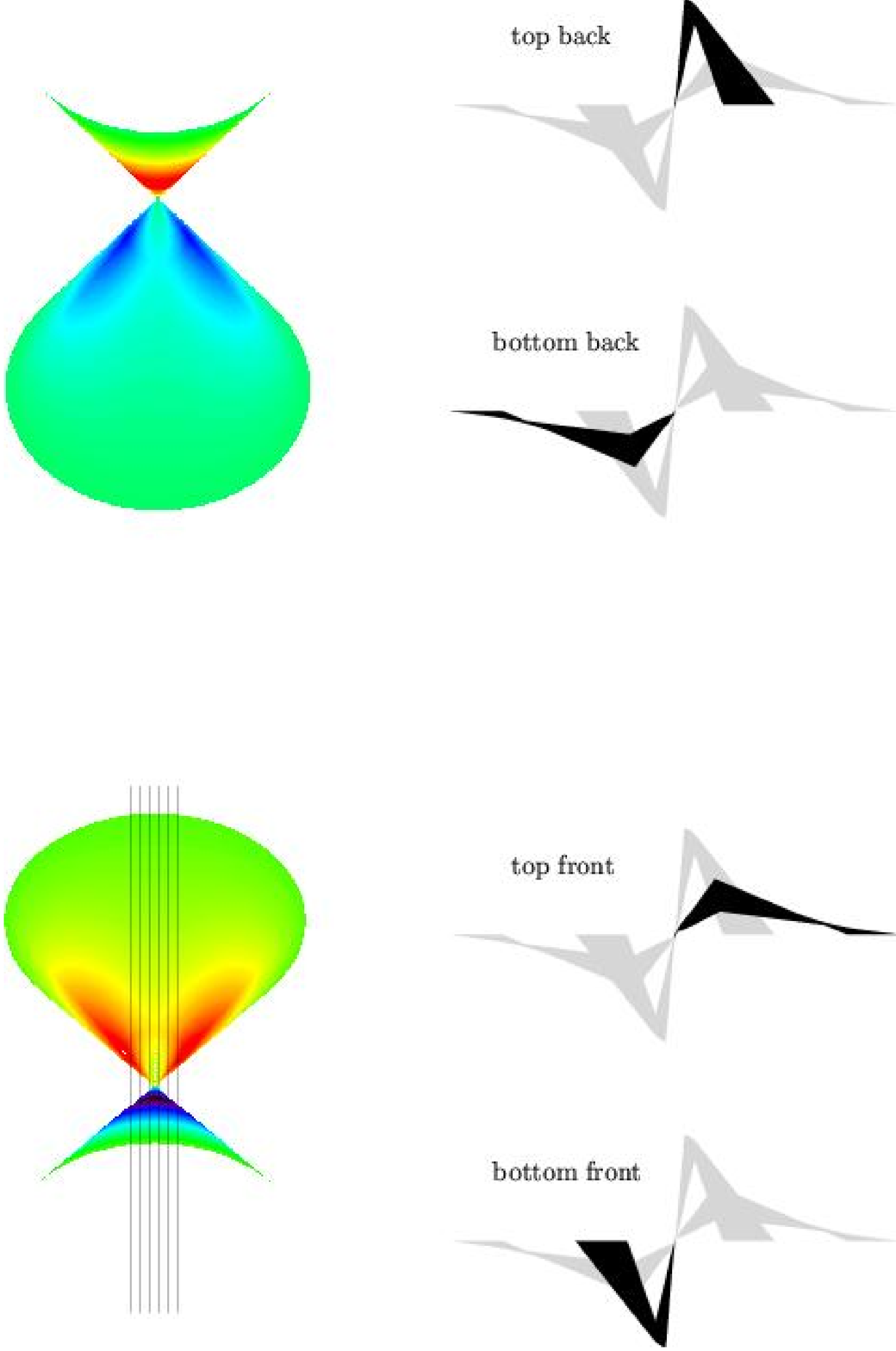]{Left: Back (upper) and front (lower) sides of the
  radial-velocity maps generated by our $v = kr$ and $ v = v_{
    max  }-k^{  '  }r  $ models. Only the outer surface of the bicone is
  shown for clarity. The position angle of the bicone axis is 60\degg\
  in the sky, so we rotate our velocity maps clockwise by 60\degg\ so that the
  projected kinematic axis is vertical. Deep purple, green, and dark red represent radial
  velocities of -700, 0, and +700 \kms, respectively. The 5 slits are placed onto the front half of the
  bicone. Right: Next to the bicones are extracted velocity plots (in black) from slit
  1, for each quadrant. The velocity plot for each quadrant is given relative to
  total extraction from slit 1 as a form of visual aid.\label{models}}

\figcaption[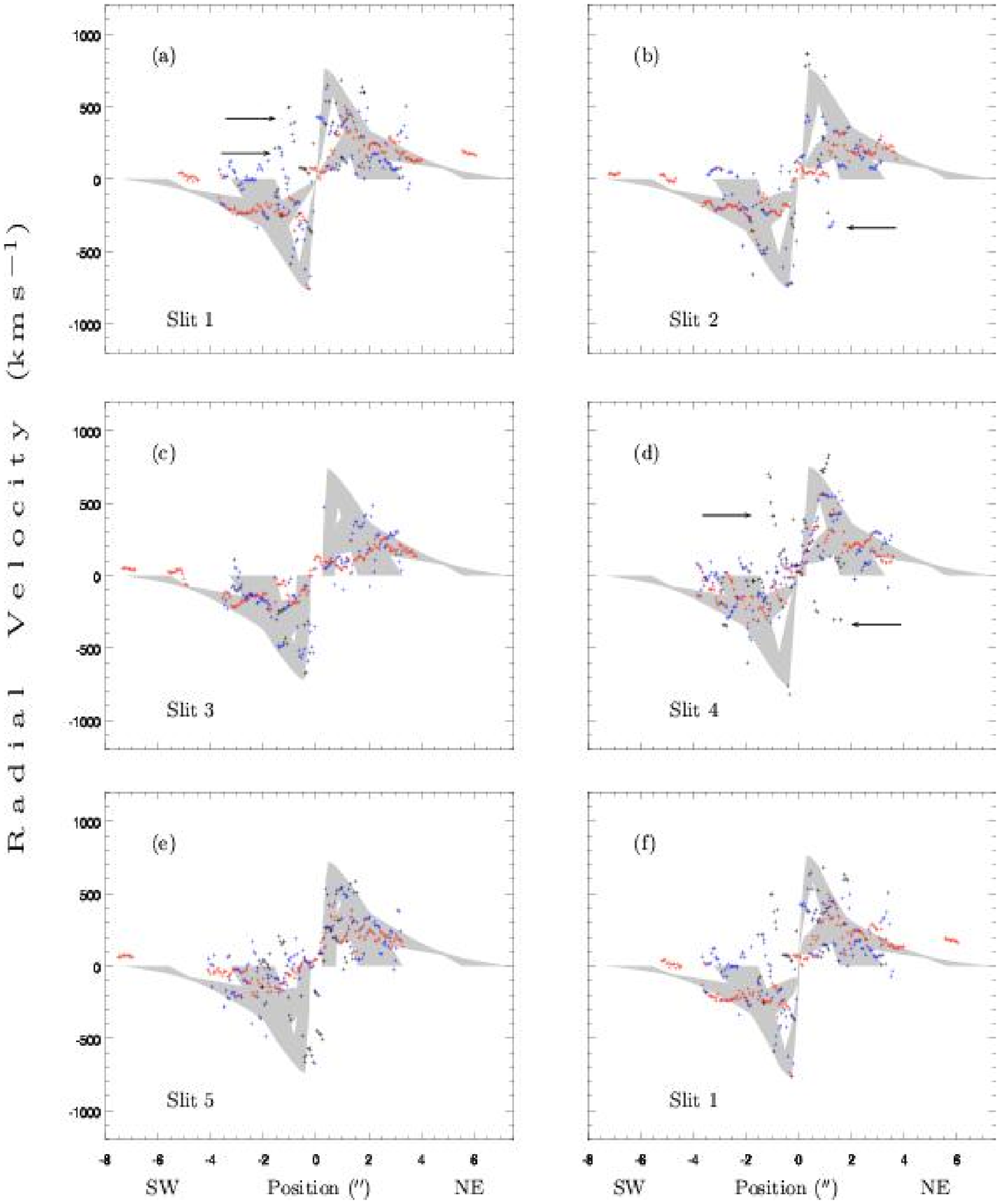]{Plots showing the model (in grey)
  and the three components in different colors. Graphs (a), (b), and (d)
  show unusual flow pattern, as indicated by the arrows. Graphs (a)
  -- (e) corresponds to slits 1 -- 5 respectively, using the $r\ law$,
  and (f) is slit 1, using the $ \sqrt r\ law  $. Positive
  position (\arcsec) is Northeast. \label{plotshade}}

\figcaption[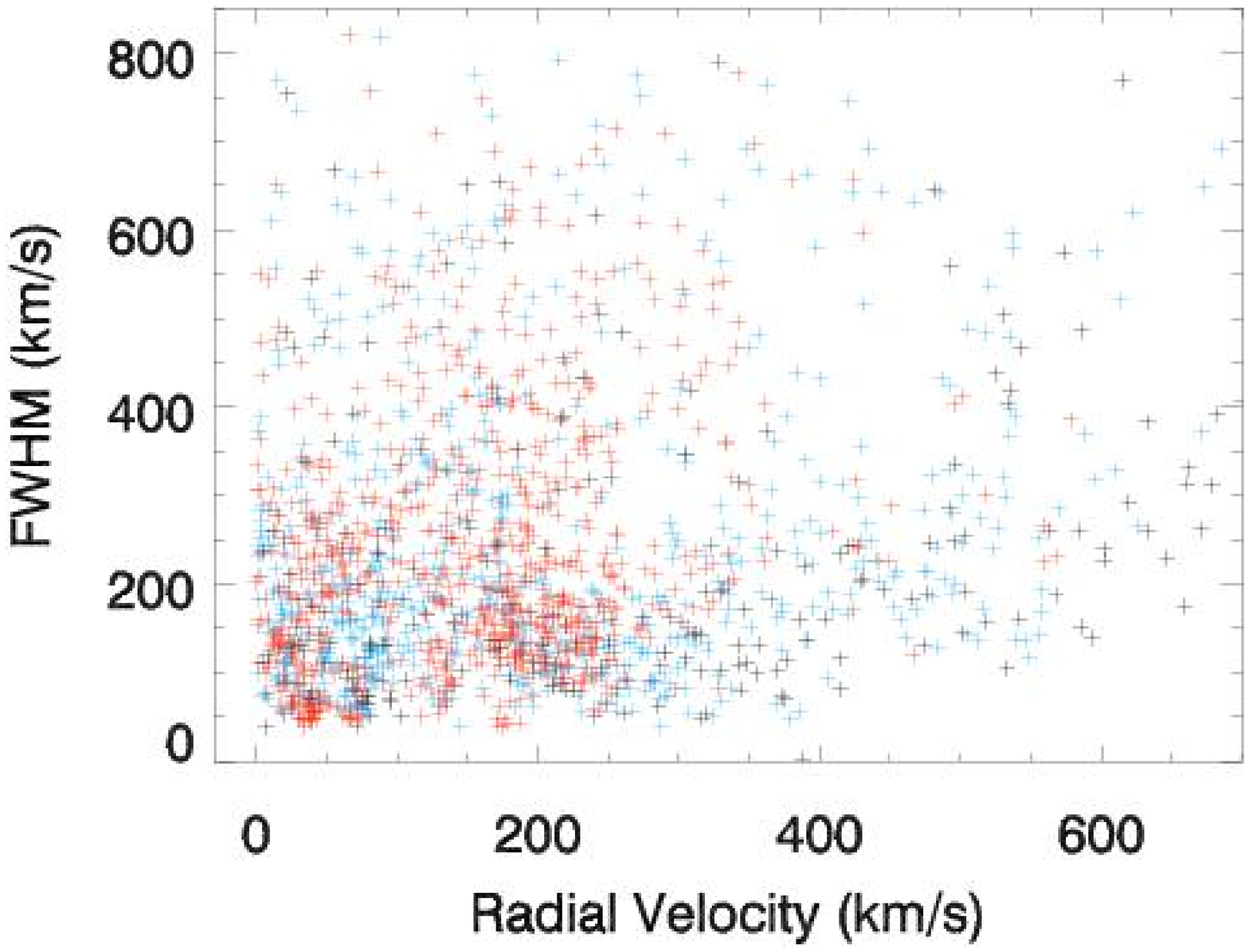]{Plot of FWHM against absolute radial
  velocity showing all five slits with high flux in red, medium flux in
  blue, and low flux in black. No obvious correlation is
  seen.\label{fwhm-vel}}

\figcaption[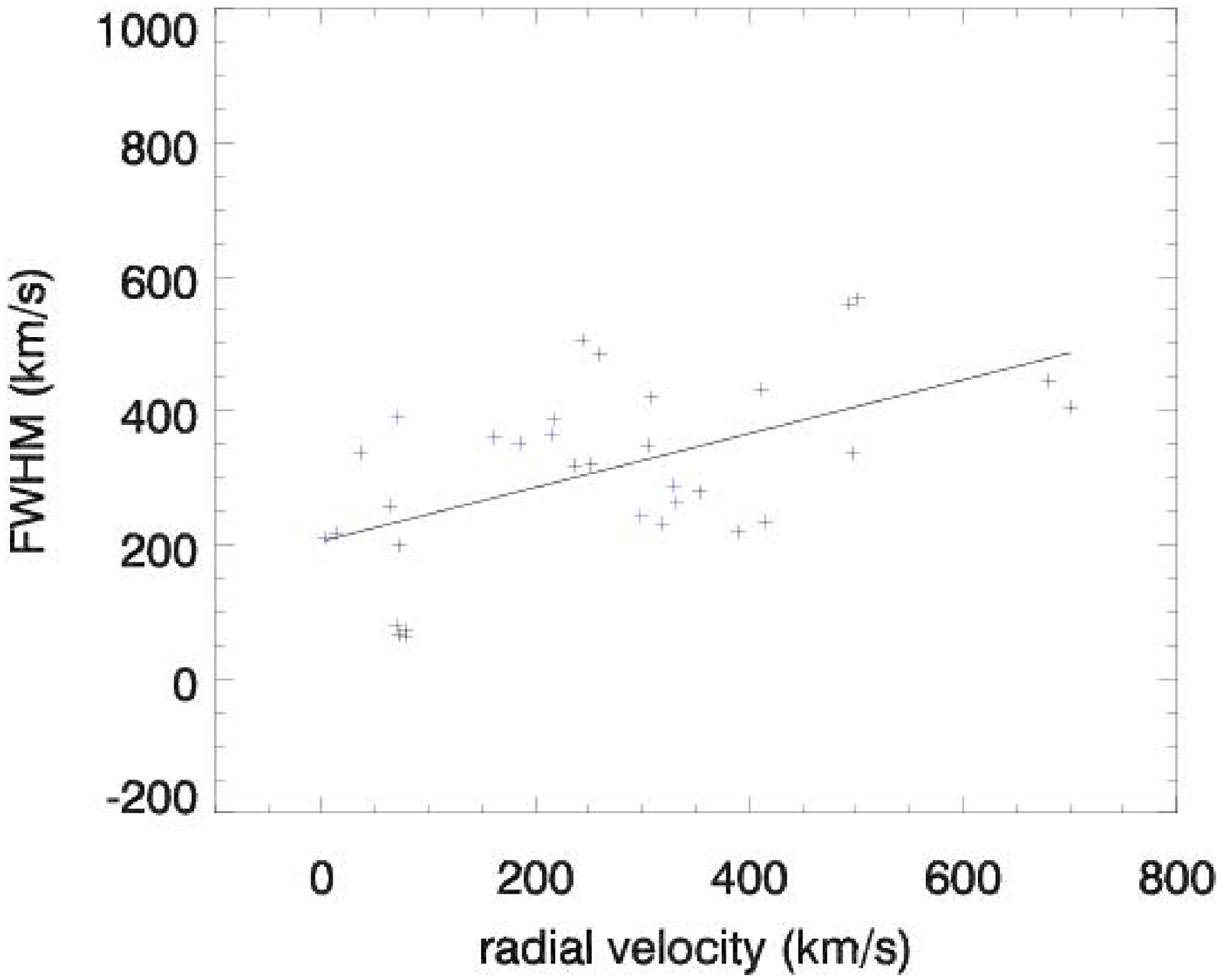]{Combined
  plots of velocity vs FWHM for knots of clouds pointed at by the arrows
  in \fig\ref{plotshade}. The straight line is a linear fit, with a
  correlation coefficient of r=0.55. \label{vel-fwhm-low-flux} }

\figcaption[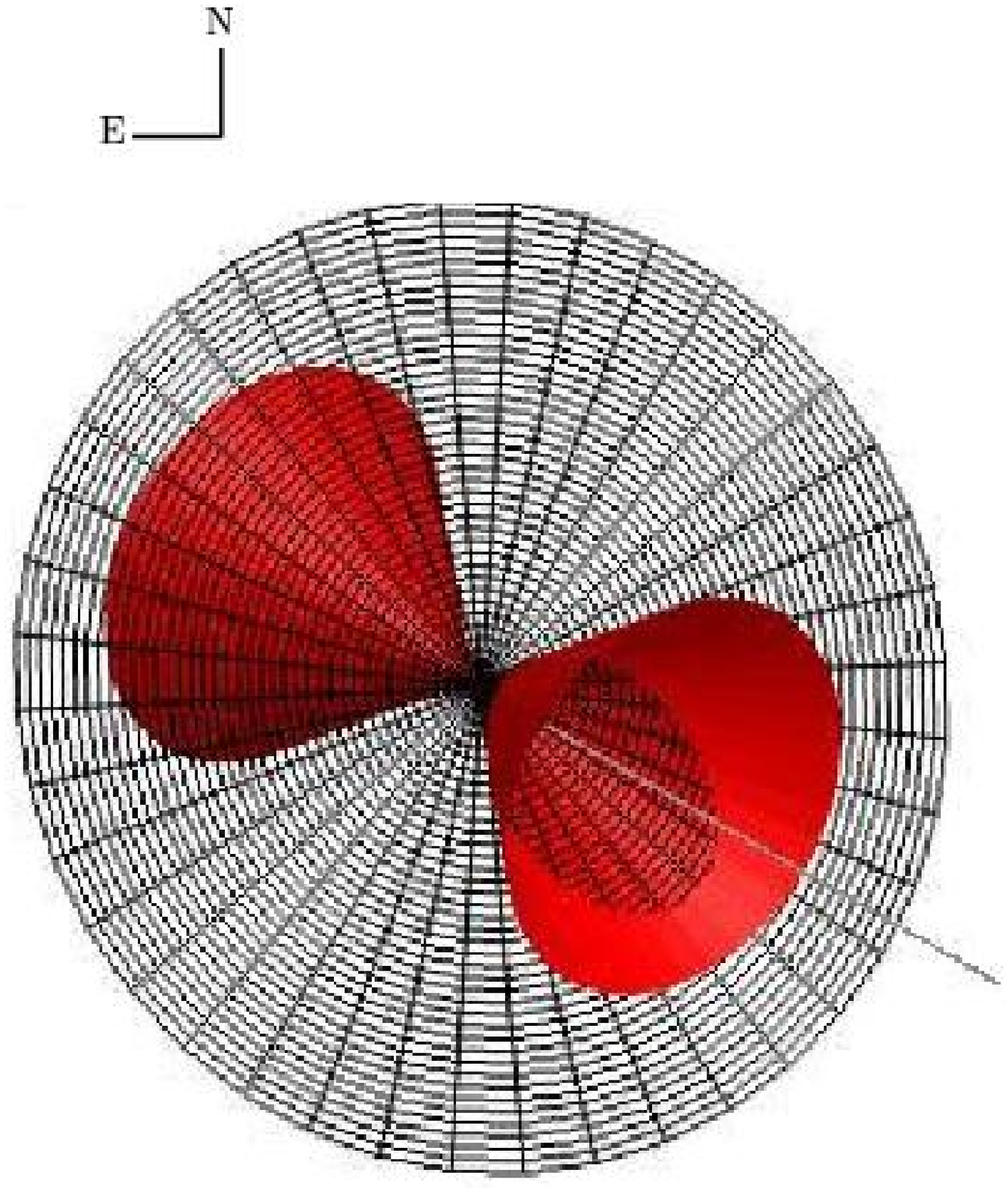]{Bicone and host
  galaxy of \galaxy\ as they should appear in
  the sky to us. The SE side of the host galaxy is the closer
  side. Also the SE bicone is closer to us. The line-of-sight is
  toward to reader.\label{bicone}}

\figcaption[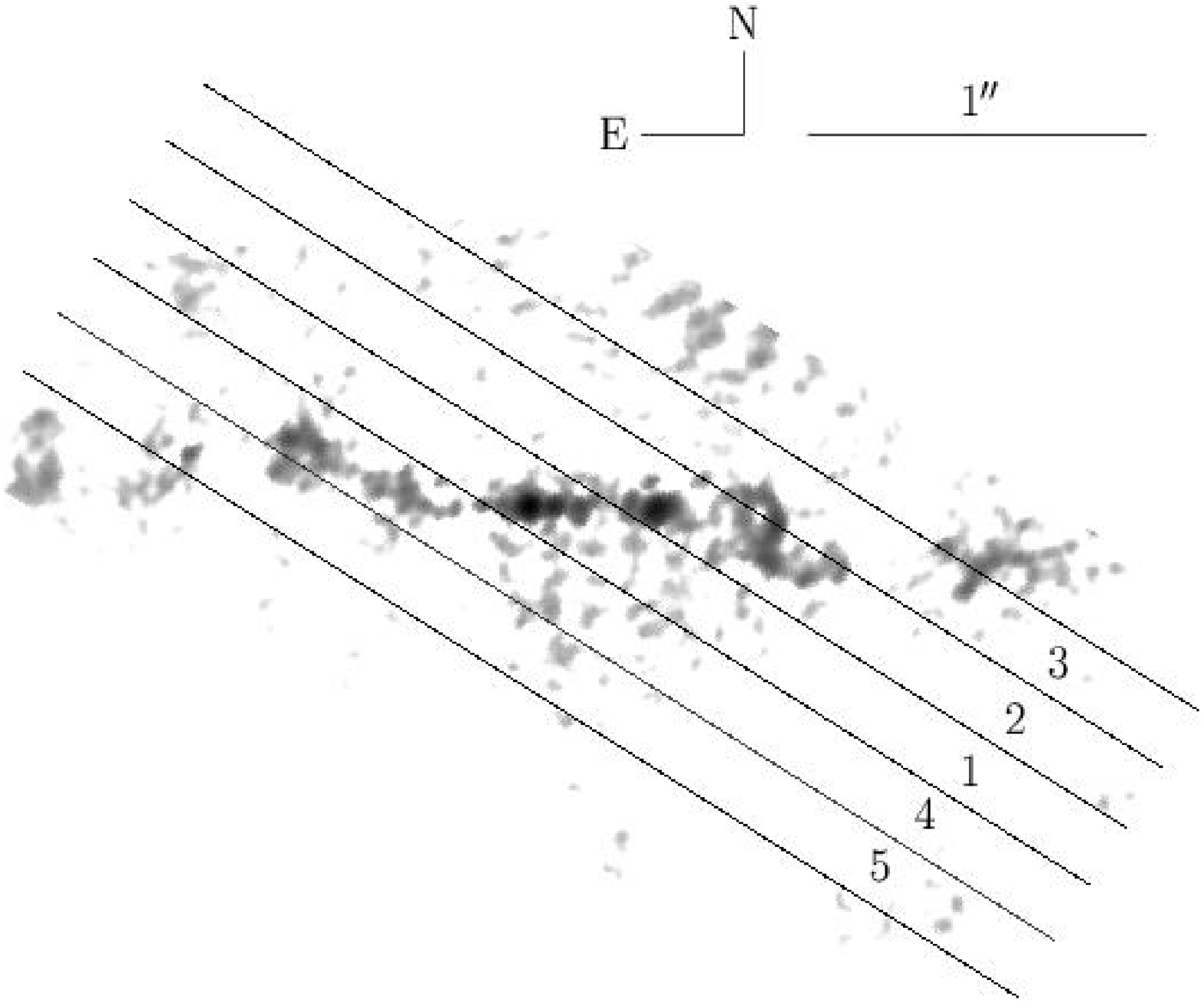]{Position of
  the slits we used for extraction from the radio image. The image was
  taken in 1996, March, by Mundell et al. The nucleus is at the second
  brightest knot in the center slit (Compare to
  \fig\ref{intensity}). \label{radio-slit-new}}

\figcaption[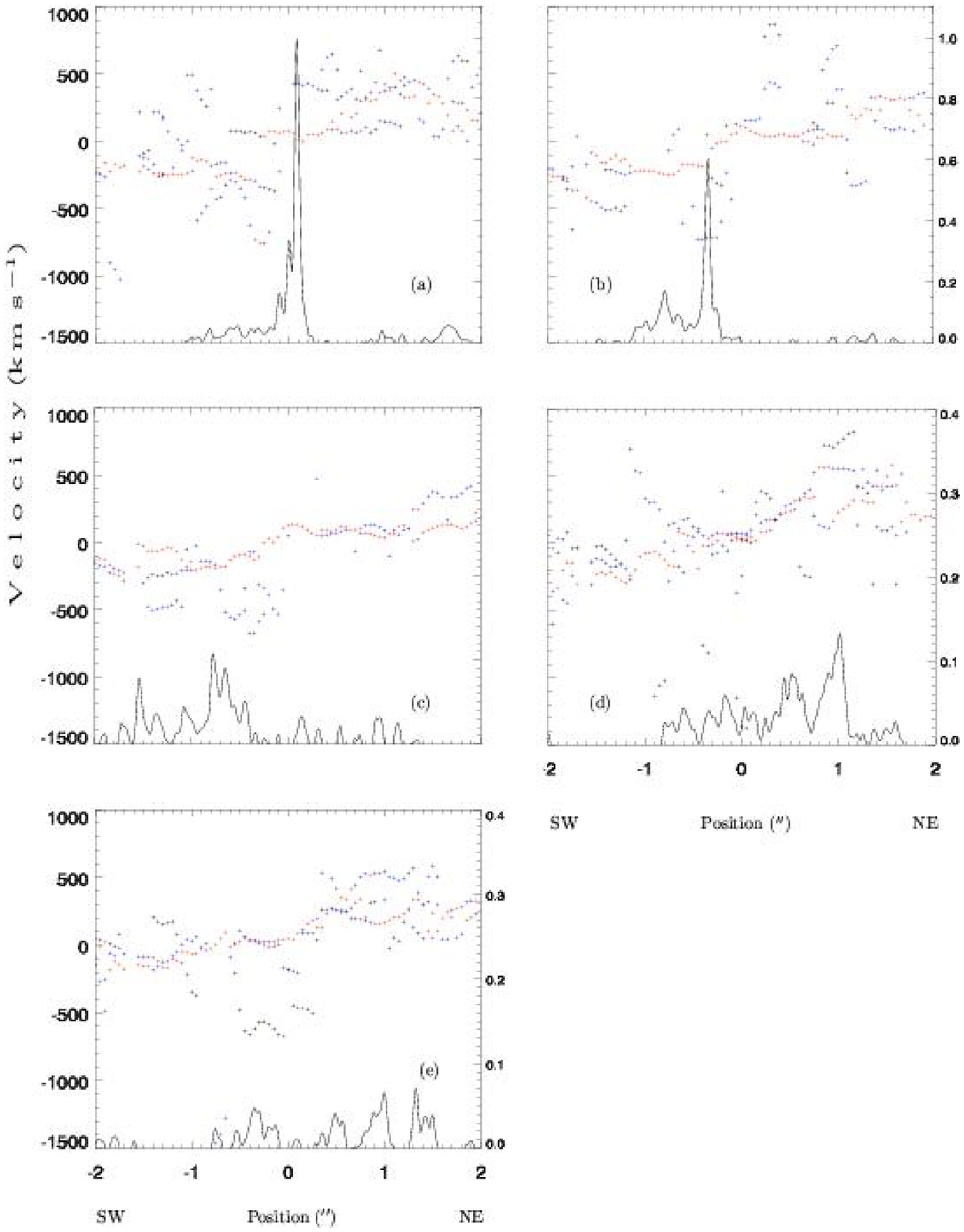]{Radio intensity (smooth
  line) and velocity
  positions (crosses) for all 5 slits, corresponding to (a) -- (e)
  respectively. The radio scale on the right side is relative to the
  bright radio knot shown in (a). \label{radio-slit-all}}

\figcaption[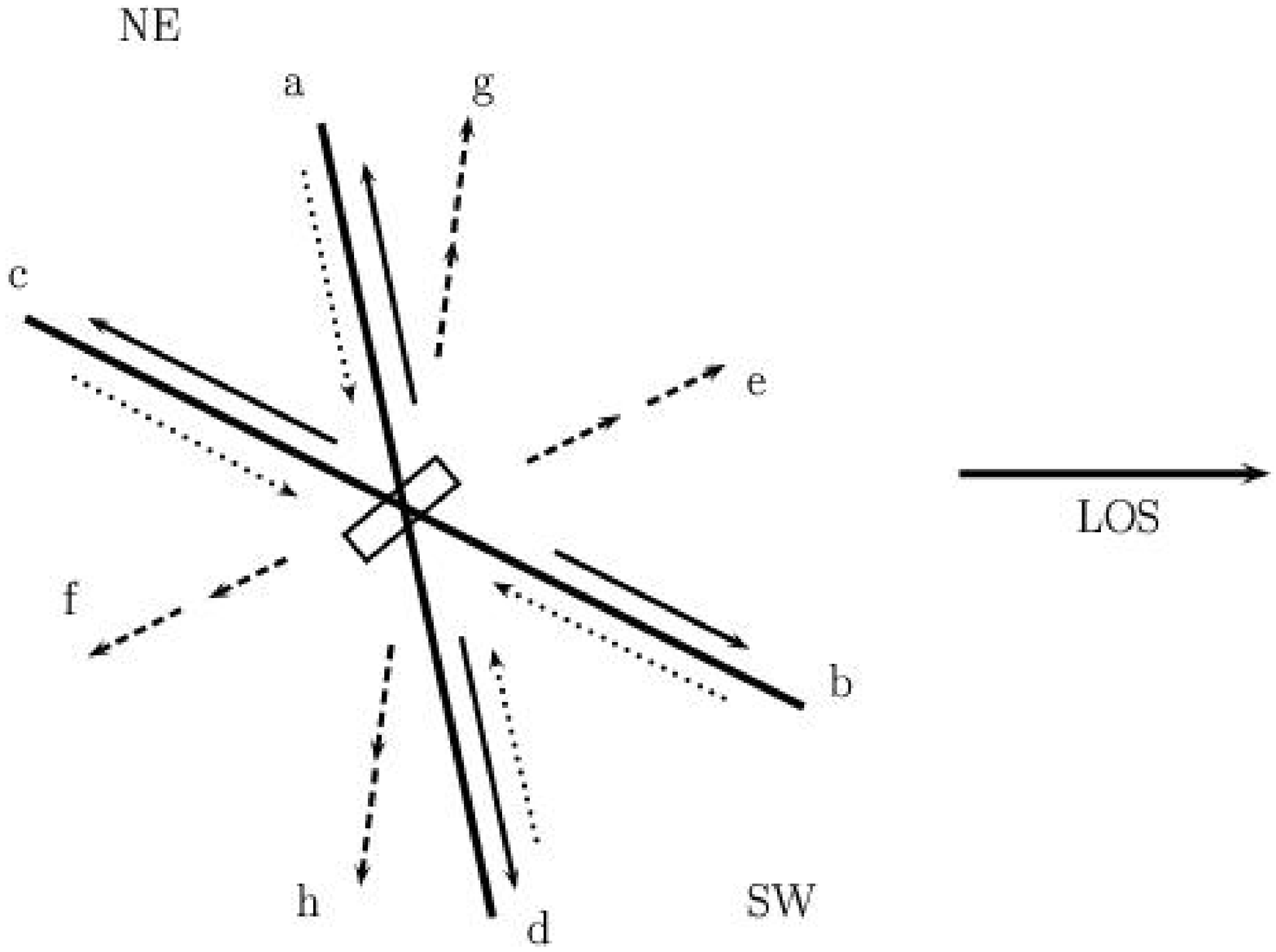]{Possible
  flow patterns for the faint rogue clouds that do not fit our model. The
  solid lines are the bicone edges. The solid arrows represent
  normal outflow corresponding to the high- and medium-flux clouds in
  the model, and the dotted and dashed arrows represent possible flow
  patterns for the rogue clouds. Note the line-of-sight (LOS) is to
  the right, and the bottom of the bicone is tilted out of the plane
  of the sky, corresponding to 45\degg\ in our model. The rectangle in
  the middle of the figure represents the torus. The largest opening
  angle is shown here (cartoon not drawn to scale).\label{backflow-cartoon}}

\clearpage

\begin{deluxetable}{@{}lll}
\tablenum{1} \tablewidth{0pt} \footnotesize
\tablecaption{\footnotesize{K{\scriptsize INEMATIC} M{\scriptsize ODEL
OF} NGC 4151 } \label{compare-ngc4151}}
\tablehead{
\colhead{Parameter} & \colhead{NGC 4151} & \colhead{NGC 4151}\\
\colhead{ } & \colhead{\footnotesize(Paper 1)} &
\colhead{\footnotesize(This Paper)}}
\startdata
\phm{sss}$z_{max}$ & \phm{ssssssss}288 pc &\phm{ssssssss}400 $ \pm
$ 16 pc\\ \phm{sss}$\theta_{inner}$ & \phm{ssssssss}20\degg
&\phm{ssssssss}15 $ \pm  $ 2\degg\\ \phm{sss}$\theta_{outer}$ &
\phm{ssssssss}36\degg &\phm{ssssssss}33 $ \pm  $ 2\degg\\ \phm{sss}$i_{axis}$ &
\phm{ssssssss}40\degg\,{\footnotesize(SW is closer)}
&\phm{ssssssss}45 $ \pm  $ 5\degg\,{\footnotesize(SW is closer)}\\
\phm{sss}$v_{max}$ & \phm{ssssssss}750\kms &\phm{ssssssss}800 $ \pm
$ 50 \kms\\ \phm{sss}$r_{t}$ & \phm{ssssssss}162 pc &\phm{ssssssss}96
$ \pm  $ 16 pc \\
\enddata
\end{deluxetable}

\clearpage

\begin{figure}
\plotone{f1.eps}
\\ Fig. 1
\end{figure}

\clearpage

\begin{figure}
\plotone{f2.eps}
\\ Fig. 2
\end{figure}

\clearpage

\begin{figure}
\plotone{f3.eps}
\\ Fig. 3
\end{figure}

\clearpage

\begin{figure}
\epsscale{0.65}
\plotone{f4.eps}
\\ Fig. 4
\end{figure}

\clearpage

\begin{figure}
\epsscale{0.75}
\plotone{f5.eps}
\\ Fig. 5
\end{figure}

\clearpage
\begin{figure}
  \plotone{f6.eps}
\\ Fig. 6
\end{figure}

\clearpage

\begin{figure}
\epsscale{1.0}
\plotone{f7.eps}
\\ Fig. 7
\end{figure}

\clearpage

\begin{figure}
\epsscale{0.8}
  \plotone{f8.eps}
\\ Fig. 8
\end{figure}

\clearpage

\begin{figure}
\epsscale{1.0}
\plotone{f9.eps}
\\ Fig. 9
\end{figure}

\clearpage

\begin{figure}
\plotone{f10.eps}
\vspace*{1in}
\\ Fig. 10
\end{figure}

\clearpage

\begin{figure}
\plotone{f11.eps}
\\ Fig. 11
\end{figure}

\clearpage

\begin{figure}
\plotone{f12.eps}
\\ Fig. 12
\end{figure}

\clearpage

\begin{figure}
\plotone{f13.eps}
\\ Fig. 13
\end{figure}

\clearpage

\begin{figure}
\plotone{f14.eps}
\\ Fig. 14
\end{figure}

\clearpage

\begin{figure}
\plotone{f15.eps}
\\ Fig. 15
\end{figure}

\end{document}